\documentclass[12pt]{article}
\usepackage{epsfig}
\usepackage{graphicx}
\begin{document}

\title{Emergent Universe Scenario in the Einstein-Gauss-Bonnet Gravity with Dilaton}
\author{B. C. Paul \thanks{bcpaul@iucaa.ernet.in} and S. Ghose \thanks{souviknbu@rediffmail.com}\\
Physics Department, North Bengal University \\
Dist : Darjeeling, PIN : 734 430, India. }

\date{}

\maketitle
\vspace{0.5in}

\abstract{We obtain cosmological solutions which admit emergent universe
(EU) scenario in the framework of Einstein Gauss-Bonnet gravity coupled
with a dilaton field in four dimensions. The coupling parameter of
the Gauss-Bonnet terms and the dilaton in the theory are determined
for obtaining an EU scenario. The corresponding dilaton potential
which admits such scenario is determined. It is found that the Gauss-
Bonnet (GB) terms coupled with a dilaton field plays an important
role in describing the dynamics of the evolution of the early as well
as the late universe. We note an interesting case where the GB term
dominates initially in the asymptotic past regime, subsequently it decreases
and thereafter its contribution in determining the dynamics of
the evolution dominates once again. We note that the Einstein's static
universe solution permitted here is unstable which the asymptotic EU
might follow. We also compare our EU model with supernova data. \\ \\ \\}

\section{Introduction: }
 Recent cosmological observations predict that the universe is passing
through an accelerated phase of expansion [1]. Although the Einstein's
General theory of Relativity (GTR) with usual matter fields
of the standard model of particle physics admits early inflation fairly
well \cite{kn:2, kn:3}, it fails to accommodate the recent accelerating phase. This
led in modern cosmology to explore a suitable theory which accommodates
observational prediction. Perhaps the present accelerating
phase of expansion may be realized either modifying the matter sector
or the gravitational sector including higher derivative terms that
are relevant in the low curvature limit. It is generally accepted that
inflation is an essential ingredient to construct cosmological models in
modern cosmology. The early inflation can be realized consistently in a
semi-classical theory of gravity \cite{kn:3, kn:4}. It may be mentioned here that
Starobinsky \cite{kn:5} obtained inflationary solution in a higher derivative
theory of gravity long before the advent of inflation actually realized.
The gravitational action of the Starobinsky model corresponds to a
theory which contains curvature squared terms in the the Einstein-
Hilbert action. However, the efficacy of inflation is known only after
the seminal work of Guth \cite{kn:3} who used phase transition mechanism
to obtain such scenario in order to resolve some of the outstanding
problems of Big bang model. The idea of inflation is very much attractive
which has been implemented in various theories leading to a
number of versions of inflation in the literature. However, the late
accelerating phase is very recently predicted from observational data
and a suitable explanation is yet to come out. To address the issue a
number of proposals came up (i) with a suitable modification of the
matter sector by incorporating exotic kind of matter fields \cite{kn:6} or (ii)
with a modification in the gravitational sector by adding curvature
squared or its inverse power terms that are effective at low curvature
limit \cite{kn:7} to the Einstein-Hilbert action. It has been shown \cite{kn:7}  that the
present accelerating phase of the universe may be obtained by adding a $\frac{\mu^4}{R}$
term to the Einstein-Hilbert action. However, Einstein-Hilbert action with the
inverse curvature correction term only is not enough as it is not free
from shortcomings. Later, it has been shown that the above shortcomings
of the theory may be removed by adding a polynomial in
Ricci scalar i.e., curvature squared term to the proposed gravitational
action \cite{kn:8}.
It was shown long ago by Zweibach \cite{kn:9} that the string corrections
due to Einstein action up to first order in the slope parameter and
fourth power of momenta should be proportional to Gauss-Bonnet
(GB) terms (where $GB = R_{\mu \nu \gamma \delta} R^{\mu \nu \gamma \delta} - 4 R_{\mu
\nu} R^{\mu \nu} + R^2$). However, it was realized subsequently that
the field redefinition theorem of 't Hooft and Veltman \cite{kn:10} may be
applicable in this case. On the Einstein's shell ($R_{\mu \nu} =
0$), an action with curvature squared term of the form $R + a
R_{\mu\nu}^2 + b R^2$ may be transformed into $ R$ itself
(neglecting higher order terms) by field redefinition :
\[
g'_{\mu \nu} = g_{\mu \nu} + a R_{\mu \nu} + g_{\mu \nu} \frac{a+2b}{2 - D} R
\]
where $D$ represents the number of dimensions. Subsequently Deser
and coworkers \cite{kn:11} have shown that on the linearized Einstein shell,
the actions $ R + \alpha' (GB)$ and $R + \alpha' R_{\mu \nu \gamma
\delta}^2$ (here $\alpha'$ is the inverse of string tension) are not
different and this result generalizes to all higher-order ghost
terms. GB terms  arise  naturally as the leading order of the
$\alpha'$ expansion of heterotic superstring theory \cite{as:1}.  It is known that GB-terms in the higher dimensions leads to ghost free
propagator. It may be mentioned here that in the framework of 4
dimensions, the GB terms do not contribute in the dynamics of
evolution. However, GB terms if coupled with a dilaton field in the
action then the combination plays an important role in the dynamics
of the evolution via the dilaton field. It has been shown also that
GB combinations coupled with scalar field plays  an important role for
avoidance of singularity in a string induced gravity \cite{kn:12}. Later, in the  braneworld scenario, it has been shown \cite{kn:13} that the naked singularities may not occur if dilaton with a Gauss-Bonnet term are considered. The issues of fine tunning in a theory
with a dilaton field and GB-terms interaction are also taken into account to 
study  in details in {\it  Ref. } \cite{kn:13, kn:14}. Recently, cosmological models with dark energy of the
universe  are probed in the Einstein-Hilbert action with GB term and it is known that the theory  accommodates the new form of energy \cite{kn:15}. It has been shown
  \cite{kn:16}  that scalar field that enters into the coupling of GB terms in the action plays an important role which may be used to explain different phases of expansion of the universe including the present accelerating phase. It is also shown that GB terms  with dilaton  admits  accelerating cosmologies  in the framework of  higher dimensions \cite{kn:17}. The Gauss-Bonnet terms in the Einstein-Hilbert action are used to obtain new black holes solution [18] and Kaluza-Klein space-times \cite{kn:19, kn:20}.Paul and Mukherjee \cite{kn:24} earlier
noted that a Gauss-Bonnet term in higher dimensions leads to
3
a 4-dimensional universe at a later epoch with many good features
considering the sign of the coupling parameter different from that one
usually gets from the low energy limit of string theory. The model also
gives a satisfactory explanation of the smallness of the effective four
dimensional cosmological constant. Recently Einstein-Hilbert action
with a combination of higher order curvature terms e.g., GB terms
including dilaton are employed to study the present acceleration of
the universe \cite{kn:16, as:2}. Therefore, the GB-theory has a rich structure
that needs to be explored. In this paper we explore emergent universe
scenario in the Einstein-Hilbert action with GB terms coupled with
dilaton field in the frame work of four dimensions. We employ the GB
term with dilaton in four dimensions to obtain an emergent universe
scenario \cite{kn:24, kn:26}.

  Earlier Harrison \cite{kn:21} obtained a cosmological solution with radiation  in the presence of a cosmological constant in a closed model of the
 universe which asymptotically approaches to Einstein static  universe but the
scenario does not exit   inflationary phase. Recently, Ellis and Maartens \cite{kn:22}
obtained similar cosmological solution considering a minimally
 coupled scalar field with a special choice of its potential where the universe
exits  from its inflationary phase followed by reheating. Subsequently it was shown by
Ellis {\it et al.} \cite{kn:27} that the potential required to obtain such scenario may be obtained naturally by a conformal transformation of Einstein-Hilbert action with $R^2$-term for a proper choice of its coupling constant. The
model incorporates an asymptotically Einstein static universe in the
past and it evolves to an accelerating universe in the framework of a
closed model of the universe. This model is usually known as emergent
universe model. The salient features of an emergent universe
scenario is that there is no time like singularity, it is ever existing and
it approaches a static universe in the infinite past ($t \rightarrow - \; \infty$). It is
interesting to construct an emergent universe model as it is capable of
solving some conceptual issues of the standard Big bang model. The
asymptotic Einstein static universe at some stage enters into the standard
Big bang phase and might have features precisely known to us.
The possibilities of an emergent universe scenario have been studied
recently in a number of theories \cite{kn:24, kn:26, kn:22, kn:23, kn:25, as:3, as:4, as:5, as:6,as:13} because it permits a
universe which is ever existing and large enough so that the spacetime
may be treated as classical entities. Recently,  Mukherjee {\it et al.} \cite{kn:24}  obtained an emergent universe (EU) scenario in a flat universe
  in the modified Starobinsky model. Subsequently,  Mukherjee {\it et al.} \cite{kn:26}
proposed a   general framework for such an EU
scenario in GR  with a mixture of matter and exotic kinds of matter
that are prescribed by an equation of state (EOS) :  $p = A \rho - B \sqrt{
\rho}$, where $A$ and $B$ are constants. The EU scenario can be realized for some possible soup
comprising primordial compositions of matter that are permitted by
the EOS. It admits existence of exotic matter and dark energy in addition
to radiation/dust \cite{kn:26}. Since recent cosmological observations
indicate that our universe is almost flat, the above emergent model
of the universe is explored in a flat universe context. Subsequently
the EU scenario proposed by Mukherjee {\it et al.} \cite {kn:24, kn:26} are explored
in the context of various theories and found that the scenario can be
realized fairly well \cite{as:4, as:5, as:6}. The purpose of the paper is to examine
EU scenario in a gravitational action with GB-terms coupled with a
dilaton field.
The plan of the paper is as follows: in sec. 2, we present the
gravitational action and set up the relevant field equations , in sec.
3, cosmological solutions with emergent universe scenario are derived.
Finally, in sec. 4, we summarize the results obtained.

\section{Action and the Field Equations:}

We consider a gravitational action  given by
\begin{equation}
I = - \int \left[ \frac{R}{2 \kappa^2}    + f(\phi) ( R_{\mu \nu \gamma \delta} R^{\mu \nu \gamma \delta} - 4 R_{\alpha \beta} R^{\alpha \beta} + R^2 ) + L_{\phi}  \right] \sqrt{- g} \;
d^{4}x
\end{equation}
where the Greek indices $\mu,  \; \nu $ represents (0, 1, 2, 3),
$ f(\phi)$ represents the coupling factor of the Gauss-Bonnet (GB) term and GB  is a  combination of squared  terms of  Riemann tensor, Ricci tensor and Ricci scalar
( $ GB = R_{\mu \nu \gamma \delta} R^{\mu \nu \gamma \delta} - 4 R_{\alpha \beta} R^{\alpha \beta} + R^2$), $g$ represents  the 4 dimensional metric,  $8\pi G = \kappa^2$ and $L_{\phi}$ represents the Lagrangian for the dilaton field. The corresponding Lagrangian for the dilaton
field  is given by
\begin{equation}
L_{\phi} = - \xi(\phi) \; \partial_{\mu} \phi \partial^{\mu} \phi - V(\phi)
\end{equation}
where $\xi (\phi)$ represents the  coupling parameter for the field
in the gravitational action and $V(\phi)$ represents  potential of
the dilaton field.

We consider a flat, homogeneous and isotropic  Robertson-Walker (RW) metric  with scale factor $a(t)$,
which is given by
\begin{equation}
ds^{2}=-dt^{2}+a^2(t)\left[dr^{2}+r^{2}(d{\theta}^{2}+sin^{2}{\theta
} \; d{\phi}^{2})\right].
\end{equation}
The action given in  (1) with the  RW metric (3)  yields the following field equations :
\begin{equation}
3 \left( \frac{\dot{a}}{a} \right)^2   = \kappa^2 \left[ \xi(\phi)  \dot{\phi}^2 + V(\phi) - 24 f'(\phi)
 \dot{\phi} \frac{\dot{a}^3}{a^3} \right],
\end{equation}
\begin{equation}
2 \frac{\ddot{a}}{a} + \left( \frac{\dot{a}}{a} \right)^2   = - \kappa^2 \left[ \xi(\phi)  \dot{\phi}^2
 - V(\phi) + 16 f'(\phi)   \dot{\phi} \frac{\dot{a} \ddot{a}}{a^2} + 8 \left(f'(\phi) \ddot{\phi} +
 f'' (\phi)  \dot{\phi}^2 \right) \frac{\dot{a}^2}{a^2} \right],
\end{equation}
once again varying the action (2.1) with respect to the dilaton field $\phi$, we get
\begin{equation}
\xi (\phi) \left[  \ddot{\phi} + 3 \frac{\dot{a}}{a} \dot{\phi} + \frac{1}{2} \frac{\xi'}{\xi} \dot{\phi}^2
 +  \frac{ V'(\phi) }{2 \xi} \right] = 12 f'(\phi)    \frac{\dot{a}^2  \ddot{a}}{a^3},
\end{equation}
where the  over dot implies derivative
w.r.t. time and prime $ (')$  represents differentiation with respect to the field $\phi$. In the above, out of the three eqs. (4)-(6), only two are independent, as eq. (2.6) can be derived from eqs. (4) and (5). 
It is evident that there are altogether  five unknowns namely, $a(t)$, $\phi$,
$V(\phi)$, $\xi(\phi)$ and $f(\phi)$ in the above field equations. In order to solve the equations  additional assumptions are can be made. Let us first assume that
\begin{equation}
f'(\phi) \dot{\phi} = \eta,
\end{equation}
 where   $\eta$ is a constant.
 The above assumption leads to a relation  $f(\phi) = \eta t (\phi) + \eta_o$, where $\eta_o$ is a  constant. The coupling parameter $f(\phi)$, therefore,
 grows with time. Consequently  the effect of the GB terms  becomes more and
 more important at late time,
 which may be useful for describing the present acceleration of the
 universe.
 Using
the constraint given by eq. (7) in eqs. (4) and (5), one gets
\begin{equation}
3 H^2   = \kappa^2 \left[\xi(\phi)  \dot{\phi}^2 + V(\phi) - 24 \eta   H^3 \right],
\end{equation}
\begin{equation}
2 \dot{H} +  3 H^2   = - \kappa^2 \left[ \xi(\phi)  \dot{\phi}^2 - V(\phi) + 16  \eta H (\dot{H} + H^2 ) \right],
\end{equation}
where $H = \frac{\dot{a}}{a}$  is the Hubble parameter.
Now eliminating $V(\phi)$ from eqs. (8) and (9), we get
\begin{equation}
\dot{H}  + \kappa^2 \left[ \xi(\phi)  \dot{\phi}^2 + 8 \eta H \dot{H} - 4 \eta H^3 ) \right] =0.
\end{equation}
The dilaton potential can be obtained from eqs. (8) and (9)
eliminating $\xi (\phi) \, \dot{\phi}^2$, the corresponding potential becomes a function of Hubble parameter which is given by
\begin{equation}
V (H) = \frac{3}{\kappa^2} H^2 + 20 \eta H^3 + \frac{1}{\kappa^2} (1 + 8 \eta \kappa^2 H) \dot{H},
\end{equation}
the above potential may be expressed as a function of $\phi$ once the Hubble parameter  ($H$) is known in terms of $\phi$, which will be determined in the next section. 
We use eqs.(8)-(9) to  determine the coupling parameters $\xi(\phi)$ 
and $f(\phi)$ in the theory for an EU cosmological solution following {\it Ref.} \cite{kn:16}. The set of eqs. (10)-(11) contain four unknowns, therefore,  two more {\it ad hoc} assumptions can be made to obtain a consistent cosmological solution. In the next section we begin with a known scale factor, which permits an emergent universe scenario. 

\section{Cosmological Solutions}
Let us consider the evolution of the scale factor of the universe in the form
\begin{equation} a(t)= a_o  \left[ A + e^{\alpha t}
\right]^{\frac{1}{\beta}}
\end{equation}
where $a_o$, $\alpha$, $\beta$ and $A$ are positive constants. It gives an EU scenario as have been obtained in {\it Refs.} \cite{kn:24, kn:26}. The Hubble parameter corresponding to eq.(12) satisfies a first order differential equation given by
\begin{equation}
\dot H = \alpha H-\beta H^2 .
\end{equation}
From eqs. (8)-(10),  the dilaton coupling and the dilaton potential are now can be determined. The field equations are highly non-linear. Consequently it is not simple to obtain a general  form of $\xi(\phi)$ and $V(\phi)$ in terms of the dilaton field. However, those parameters  may be determined in terms of the Hubble parameter following {\it Ref.} \cite{kn:16}, which are 
\begin{equation}
\xi(H)=\frac {1}{\kappa^2 \dot\phi^2}\left[4\eta\kappa^2 \left(1+2\beta \right)H^3 + \left(\beta-8\kappa^2 
\alpha \eta \right)H^2-\alpha H \right] ,
\end{equation}
\begin{equation}
 V(H)=\frac{3}{\kappa^2}H^2+20\eta H^3+\frac{1}{\kappa^2}\left(1+8\kappa^2 \eta H \right)
 \left(\alpha H -\beta H^2 \right) .
\end{equation}

For simplicity, as a special case let us consider  $\beta = 8 \eta \kappa^2 \alpha$ in the above.the corresponding
dilaton field potential can be expressed in terms of the field. There is
another freedom to assume here as the number of unknowns are one
more than the number of relevant equations. Therefore, we look for
emergent universe scenario for different behaviors of the dilaton field
in the next subsection.

\subsection{Case I :}  For an increasing dilaton field $\phi=\phi _o e^{\alpha t} $, with 
the GB coupling terms $f(\phi)=\frac{\eta}
{\alpha}\;ln \; \phi\;$, dilaton coupling is given by
\begin{equation}
\xi(\phi)=\frac{1}{2 \kappa^2 \phi (1+ \beta \phi)^3}  \left(\beta  \phi^2 -  4 \beta \phi -2 \right)
\end {equation}
We note the following: 
$(i)$  $\xi(\phi)\rightarrow\infty $ when $\dot a\rightarrow 0$ i.e. $ H\rightarrow 0$ and
$(ii)$ $\xi(\phi)\rightarrow 0 $  at two points (a) for $H_1=\frac{\alpha}{\beta}$ and (b) for $H_2= \frac {2\alpha^2}{\beta (1+2\beta)}.$ We plot the  variation of the dilaton coupling  ($\xi$) with $\phi$ as shown in the fig.$1$. It is evident that initially the dilaton coupling begins with negative value (phantom like property) but in course of its evolution $\xi(\phi)$ becomes positive and almost constant after attaining a peak. Thus in this model to
begin with one can start with a field having negative kinetic energy,
an interesting field which behaves like phantom \cite{as:7}, now-a-days it
is considered as one of the candidate of dark matter and avoids the
singularity. The potential is given by
\begin{equation}
V(\phi) = \frac{\alpha^2}{\kappa^2 (1 + \beta \phi)^3}  \left( \frac{11}{2} \beta \phi^3 + (2 \beta +3) \phi^2 + \phi \right).
\end{equation}
The Hubble parameter in this case is related to field as 
\begin{equation}
H = \frac{\alpha \phi}{1+\beta \phi}.
\end{equation}
 We determine  $ A=1$ , $ \phi_o = \frac{1}{\beta} $ and 
the corresponding evolution of the scale factor becomes
\begin{equation}
a =a_o \left[1+ e^{\alpha t} \right]^{\frac{1}{\beta}}.
\end{equation}

\begin{figure}
\begin{center}
\includegraphics[width=5cm,height=5cm]{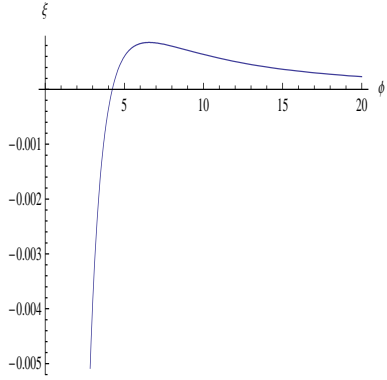}
\caption[$\xi( \phi)$ Vs. $\phi$ curve for $\kappa=1$ and $\beta=2$]{$\xi( \phi)$ Vs. $\phi$ curve for $\kappa=1$ and $\beta=2$}
\end{center}
\end{figure}

\subsection{Case II :}   For a decreasing dilaton field $\phi=\phi _o e^{-\alpha t} $, with the GB coupling term $f(\phi)=- \frac{\eta}
{\alpha} ln \, \phi $, the dilaton coupling is given by
\begin{equation}
\xi(\phi)=\frac{1}{2\kappa^2  \phi^2 \,(\beta + \phi)^3 }   \left(\beta - 4 \beta \phi - 2 \phi^2  \right)
\end{equation}
In fig. 2 we plot variation of dilaton coupling $ \xi(\phi) $ with $\phi$. It is evident that $\xi (\phi)$ begins with a negative value then attains a minimum thereafter it increases as $\phi$ decreases. The potential for the dilaton field is obtained as
\begin{equation}
V(\phi) = \frac{\alpha^2}{\kappa^2 (\beta + \phi)^3)} \left( \frac{11}{2} \beta  + (2 \beta +3) \phi + \phi^2 \right).
\end{equation}
The Hubble parameter is given in terms of the dilaton as
\begin{equation}
H = \frac{\alpha}{\beta + \phi}
\end{equation}
the corresponding scale factor is
\begin{equation}
a =a_o \left[\phi _o+\beta  e^{\alpha t} \right]^{\frac{1}{\beta}}.
\end{equation}

\begin{figure}
\begin{center}
\includegraphics[width=5cm,height=5cm]{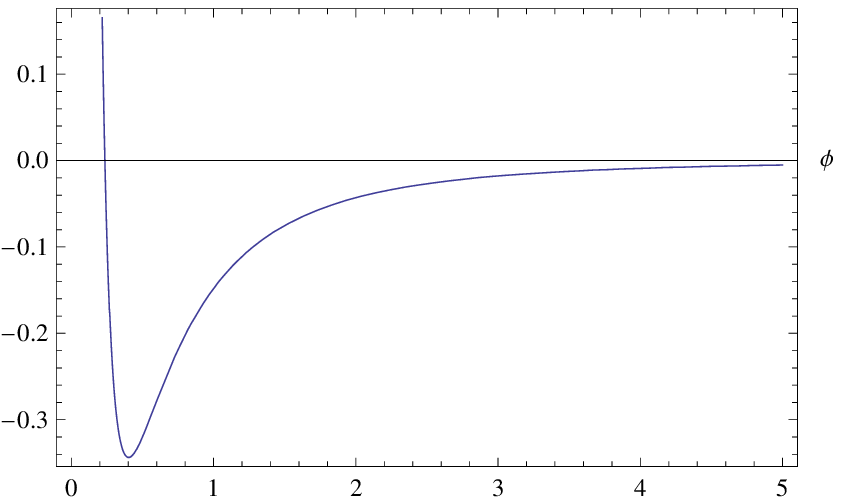}
\caption[$\xi(\phi)$ vs. $\phi$ curve for $\kappa=1$ and $\beta=2$]{$\xi(\phi)$ vs. $\phi$ curve for $\kappa=1$ and $\beta=2$}
\end{center}
\end{figure}

\subsection{Case III :} For a slowly varying field $ \phi=\frac{1}{\alpha} \, \ln \, t $, with $ f(\phi)=\eta \, e^{\alpha
\phi}$, similar to that one expects in the string theory framework [34], the
dilaton coupling is given by
\begin{equation}
\xi(\phi)=\frac{1}{\kappa^2}\left[\frac{\beta(1+2\beta)}{2\alpha}H^3-\alpha H\right]\left(ln \frac{H}{\alpha-\beta H} \right)^2,
\end{equation}
and the dilaton potential is
\begin{equation}
V(H) = \frac{1}{\kappa^2}  \left( \alpha H + 3 H^2 + \frac{\beta}{\alpha} \left( \frac{5}{2} - \beta \right) H^3 \right),
\end{equation}
where  the Hubble parameter (H) is given by 

\begin{equation}
H=\frac{\alpha e^{\alpha e^{\alpha \phi}}}{1+\beta e^{\alpha e^{\alpha \phi}}}.
\end{equation}
Fig. 3 shows variation of $\xi(\phi)$ with $\phi$ for $\beta= 5$. Here the dilaton coupling decreases as the dilaton increases to begin with from a negative value initially, attains a minimum, thereafter, it increases sharply. GB terms become important at late time in this case. The potential becomes flat as $\phi \rightarrow \infty$ which is shown in fig. 4.
\begin{figure}
\begin{center}
\includegraphics[width=5cm,height=5cm]{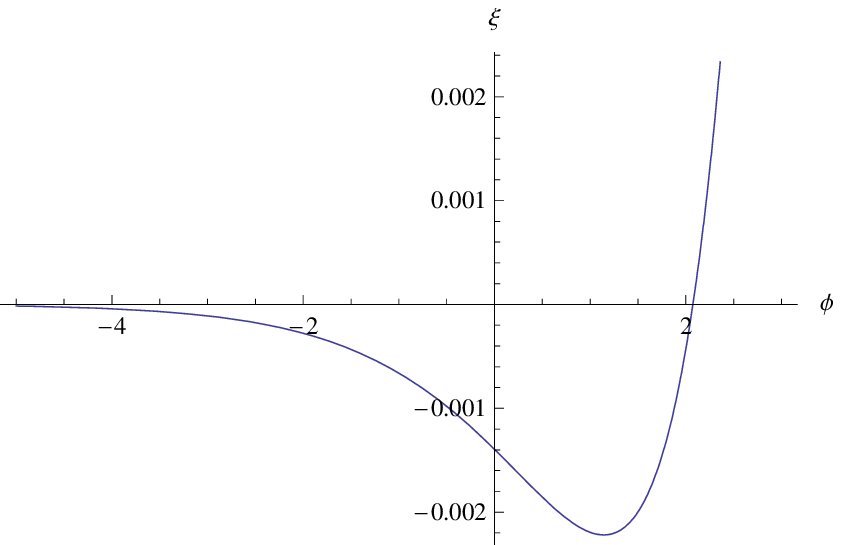}
\caption[$\xi(\phi)$ Vs. $\phi$ curve for  $\beta=5$]{$\xi(\phi)$ Vs. $\phi$ curve for  $\beta=5$}

\includegraphics[width=5cm,height=5cm]{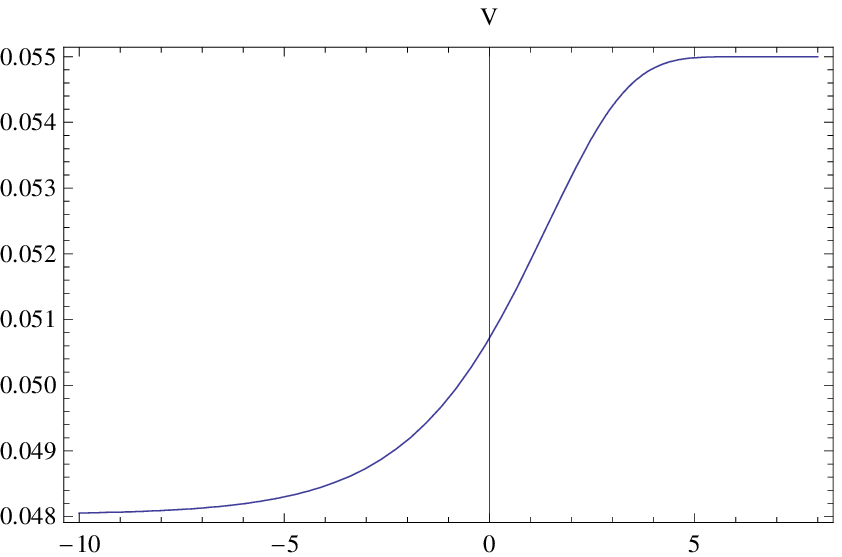}
\caption[$V(\phi)$ vs. $\phi$ curve for $\alpha = 0.5$ and  $\beta= 5$]{$V(\phi)$ vs. $\phi$ curve for $\alpha = 0.5$ and  $\beta= 5$}
\end{center}
\end{figure}

\subsection{Case IV :} For a dilaton field  $\phi = H,$ considering variation of the coupling of GB term as $f=\eta \, t$.  We obtain :
\begin{equation}
\xi(\phi)=\frac{1}{\kappa^2 \phi^2 (\alpha-\beta \phi)^2}\left[\frac{\beta(1+2\beta)}{2\alpha}\phi^2-\alpha \right]
\end{equation}
with dilaton potential given by
\begin{equation}
V(\phi)=\frac{1}{\kappa^2}\left(\alpha \phi +3\phi^2+\frac{\beta(5-2\beta)}{2\alpha}\phi^3\right)
\end{equation}

The fig. 5 shows the variation of  $\xi(\phi)$ vrs. $\phi$, for $\beta = 5$. We note that the coupling parameter becomes undetermined when $ \phi = \frac{\alpha}{\beta}$. In  this case it is necessary to begin with an initial field which greater than the above limiting value.  In that case the coupling parameter always remains positive definite and it decreases from a large value to zero. Here the GB terms do not contribute at late time as the coupling $\xi \rightarrow 0$.  The corresponding dilaton potential required for the EU model is shown in fig. 6. It has a minimum which is negative definite.The dilaton potential for $\beta = \frac{5}{2}$, becomes
\begin{equation}
V(\phi)=\frac{3}{\kappa^2}\left(\phi+\frac{\alpha}{6}\right)^2-
\frac{\alpha^2}{12\kappa^2}.
\end{equation}
In this case the universe evolves as
\begin{equation}
a = a_o \left( 1 + \frac{5}{2} \, e^{\alpha \, t} \right)^{\frac{2}{5}}.
\end{equation}

\begin{figure}
\begin{center}
\includegraphics[width=5cm,height=5cm]{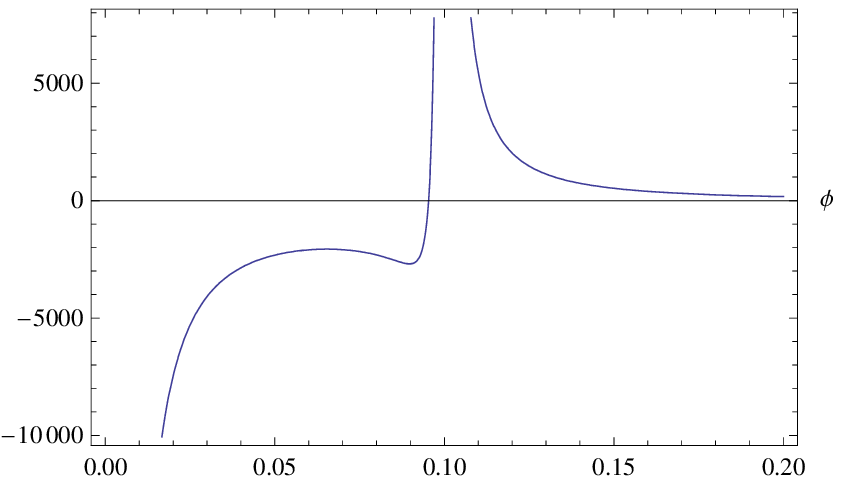}
\caption[$\xi(\phi)$ vs. $\phi$ curve for $\alpha= 0.5$ and $\beta=5$]{$\xi(\phi)$ vs. $\phi$ curve for $\alpha= 0.5$ and $\beta=5$}
\quad
\includegraphics[width=5cm,height=5cm]{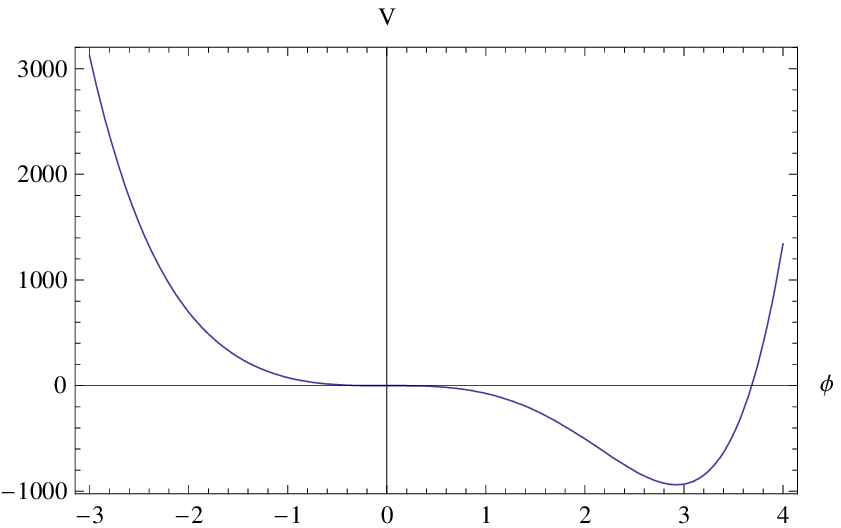}
\caption[$V(\phi)$ vs. $\phi$ curve for $\alpha = 0.5$ and $\beta= 5$]{$V(\phi)$ vs. $\phi$ curve for $\alpha = 0.5$ and $\beta= 5$}
\end{center}
\end{figure}

\subsection{Case V :} We now consider a special case $f (\phi ) = \eta_o + \eta \phi^2$ to obtain EU scenario. In this case the dilaton evolve as $\phi = \pm \sqrt{t} \;$ with the coupling parameter
\begin{equation}
\xi (\phi) = \frac{4 \phi^2}{\kappa^2} \left( - \alpha H + \frac{\beta}{2 \alpha} (1 + 2 \beta) H^3 \right).
\end{equation}
 The dilaton field potential is
\begin{equation}
V(\phi) = \frac{1}{\kappa^2} \left( \alpha H + 3 H^2 + \frac{\beta}{\alpha} \left(\frac{5}{2} - \beta \right) H^3 \right)
\end{equation}
where $ H = \frac{ \alpha e^{\alpha \phi^2}}{1+ \beta  e^{\alpha \phi^2}} $.

Fig. 7 shows the variation of the dilaton coupling parameter with
field, which is interesting. The variation of the potential is also shown
in fig. 8, which has two flat regions one in the early era and the
other in the late era respectively. This is new and interesting as it
permits both early inflation and late acceleration. In {\it Ref.} \cite{kn:24} it
was shown that a particle creation may occur during a phase when
the Hubble parameter varies slowly. In this case the GB coupling
parameter $f(\phi) \rightarrow \pm \infty$ as $ t 
\rightarrow \pm \infty$. The GB combination might have
dominated in the early epoch of evolution which eventually decreases
at later epoch (for $\eta_o=0$)  corresponding to a minimum of $f(\phi)$ (say at $t = 0$) thereafter it increases which is shown in fig. 9. Thus GB
terms play an important role in the early era which is subsequently
important once again at late era contributing to the dark energy \cite{kn:8}. 
\begin{figure}
\begin{center}
\includegraphics[width=5cm,height=5cm]{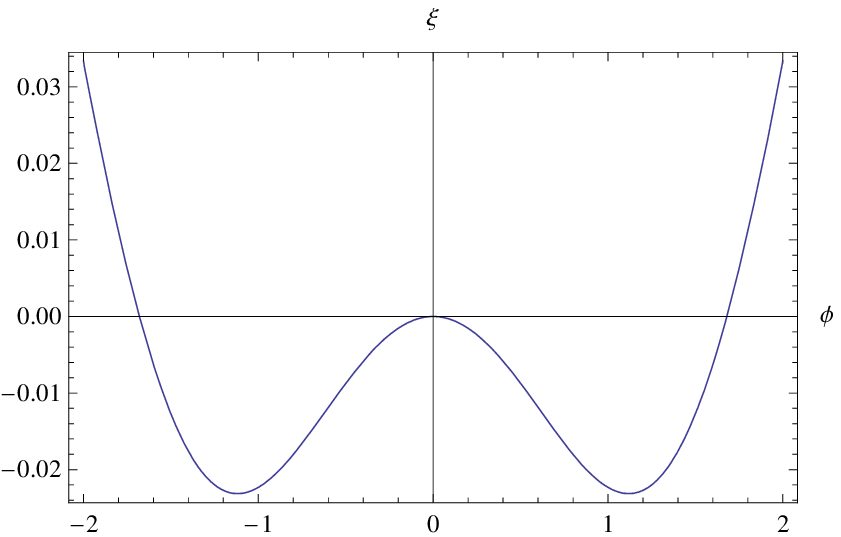}
\caption[$\xi(\phi)$ vs. $\phi$ curve for $\beta= 5$]{$\xi(\phi)$ vs. $\phi$ curve for $\beta= 5$}
\includegraphics[width=5cm,height=5cm]{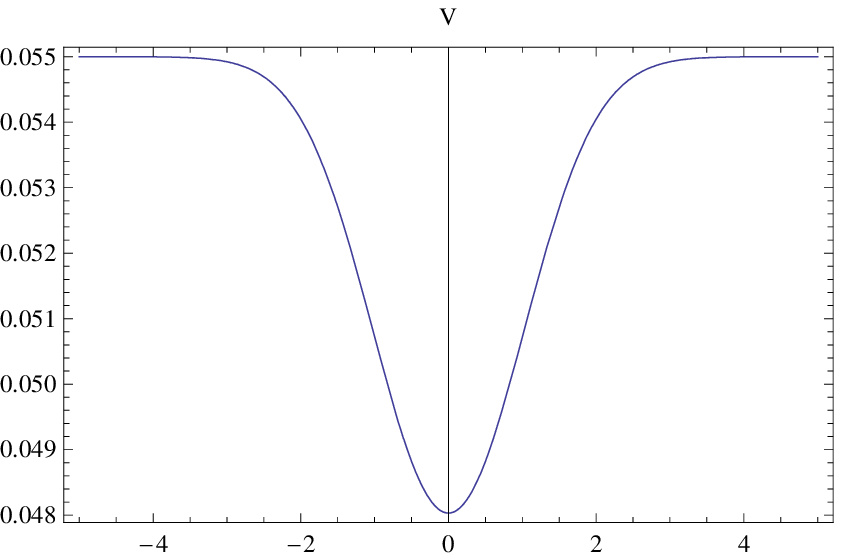}
\caption[$V(\phi)$ vs. $\phi$ curve for  $\beta=5$]{$V(\phi)$ vs. $\phi$ curve for  $\beta=5$}
\includegraphics[width=5cm,height=5cm]{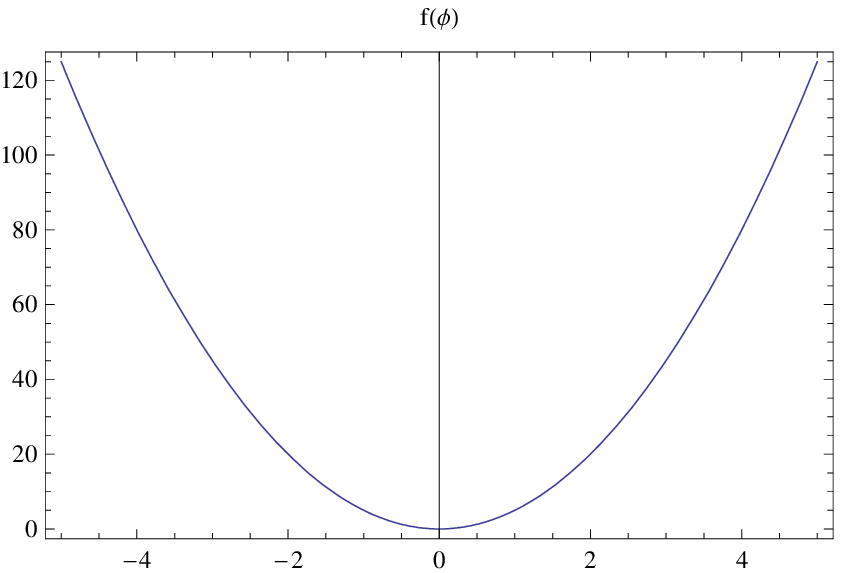}
\end{center}
\caption[$f(\phi)$ vs. $\phi$ curve for  $\eta_o= 0$ and $\eta = 5$]{$f(\phi)$ vs. $\phi$ curve for  $\eta_o= 0$ and $\eta = 5$}
\end{figure}

\section{Stability of Einstein's Static Universe}
 The EU scenario is characterized by the scale factor $a(t)= (a_0 +\beta e^{\alpha t})^{1/\beta}$ as given in (12). It may be pointed out here that the universe (i) begins
with singularity if $\beta< 0$, (ii) begins with singularity and asymptotically
approaches Einstein static (ES) universe at late time if $\alpha < 0$ and
$\beta\neq 0$, and (iii) spends infinite time near the Einstein static universe
but pulls away and ends in an infinite inflating epoch if $\alpha > 0$ and
$\beta > 0$. It is pointed out that the first two cases lead to unstable
solution \cite{as:8}.
 It may be pointed out here that: (i) The existence of the ES universe in fourth order theories of gravity and stability of the ES has been studied in \cite{as:10}. (ii) Stability of ES in $f(R)$ gravity has also been considered in \cite{as:11}. (iii) It was also noted that ES universes are unstable in generic $f(R)$ models \cite{as:12}.
 To analyze the stability of Einstein's Static (ES) universe (which
is an asymptotic past solution) \cite{as:9} in the theory we begin with a pair
of differential equations in $a$ and $H$ which are given by
\begin{equation}
\dot{a}= a H.
\end{equation}
\begin{equation}
\dot{H}=\alpha H - \beta H^2, 
\end{equation}
The above equations form an autonomous system which can be analyzed
by the standard technique. The Einstein static universe solution
corresponds to the critical point of the system $(a_0, 0)$. The ES universe in this case is unstable for $\alpha > 0$.
It may be pointed out here that the stability of the Einstein Static universe under inhomogeneous perturbations has recently been studied in {\it Refs.} \cite{as:14}.

\section{Distance modulus curve:}

We now probe late universe in the Emergent Universe scenario taking
into account the observational results available from supernova. The
distance modulus is $\mu = 5 \log d_L + 25$ where $d_L$ is the luminosity distance (in the unit of mega parsecs), given by
 \begin{equation}
 d_L = r_1 (1+z) a(t_0)
 \end{equation}
 where $r_1$ is given by :
  \begin{equation}
 \int^{r_1}_{0} \frac{dr}{\sqrt{1-kr^2})}=  \int^{t_0}_{t_1} \frac{dt}{a(t)}
 \end{equation}
 We consider the scale factor for a EU scenario as was given in (12),
   \begin{equation}
    a(t)=a_0(\sigma+e^{\alpha t})^{\omega},
 \end{equation}
 where $a_0,\sigma$ and $\omega$ are constants. At late time the exponential term dominates and one can write
 \begin{equation}
 a(t)\sim a_0e^{\alpha \omega t}
 \end{equation}
 Since we are considering a flat universe, $k=0$, eq. (36) yields
   \begin{equation}
 r_1= \int^{t_0}_{t_1} \frac{dt}{a(t)}
 \end{equation}
Using the scale factor (37) one obtains an expression for dL which is
given by
 \begin{equation}
 d_L= \frac{z(1+z)}{H_0}
 \end{equation}
 Note that the final expression for $d_L$ does not depend on $a_0, \sigma$ and $\omega$. It is now possible to determine $\mu(z)$ numerically for an Emergent
Universe scenario at different values of redshift parameters $(z)$. The
observed values of $\mu(z)$ at different $z$ parameters \cite{kn:1} along with that
obtained from the present theory are given in Table : 1. 
\begin{figure}
\begin{center}
\includegraphics[width=6cm,height=6cm]{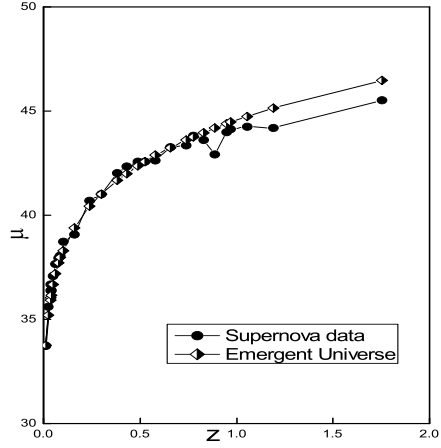} 
\caption[$\mu$ vs. $z$ curve for Emergent universe and supernova data.]{$\mu$ vs. $z$ curve for Emergent universe and supernova data.}
\end{center}
\end{figure}
\begin{center}
\begin{tabular}{|c|c|c|} 
\hline 
\emph{$z$} & \emph{Supernova $\mu$} & \emph{EU $\mu$} \\ \hline
0.038 &	36.67 &	36.0438240 \\
0.014 &	33.73 &	33.7609468 \\
0.026 &	35.62 &	35.1945228 \\
0.036 &	36.39 &	35.9222306 \\
0.040 &	36.38 &	36.1593860 \\
0.050 &	37.08 &	36.6647158 \\
0.063 &	37.67 &	37.1932883 \\
0.079 &	37.94 &	37.7171620 \\
0.088 &	38.07 &	37.9694771 \\
0.101 &	38.73 &	38.2944627 \\
0.160 &	39.08 &	39.4068091 \\
0.240 &	40.68 &	40.4320839 \\
0.300 &	41.01 &	41.0192423 \\
0.380 &	42.02 &	41.6622327 \\
0.430 &	42.33 &	42.0079418 \\
0.490 &	42.58 &	42.3808310 \\
0.526 &	42.56 &	42.5866208 \\
0.581 &	42.63 &	42.8794593 \\
0.657 &	43.27 &	43.2483587 \\
0.740 &	43.35 &	43.6128241 \\
0.778 &	43.81	& 43.7684760 \\
0.828 &	43.61	& 43.9639519 \\
0.886 &	42.91	& 44.1787963 \\
0.949 &	43.99	& 44.3993095 \\
0.970 &	44.13	& 44.4701090 \\
1.056 &	44.25	& 44.7473544 \\
1.190 &	44.19	& 45.1438747 \\
1.755 &	45.53	& 46.4859129 \\ \hline
\end{tabular}
\end{center}

\section{Discussion}
We obtain emergent universe scenario in a modified theory of gravity
with Gauss-Bonnet term coupled to a dilaton field in four dimensions.
We look for an emergent universe (EU) scenario here in a spatially flat
universe. In the EU model, the universe was in an almost static state
in an infinite past which eventually evolves into an inflationary stage
later on. The Einstein static universe permitted here is found unstable.
We note that emergent universe scenario obtained in {\it Refs.} \cite{kn:24, kn:26} can be implemented in a modified theory of gravity with GB terms
quite successfully with a suitable dilaton and GB coupling parameters.
The parameters are determined here and the corresponding models are presented in cases I to
V in section 3. The result obtained here once again supports the view
that the GB combination might have played a crucial role for driving
both early inflation and late acceleration. A new solution is noted here
in which the coupling parameter of the GB terms becomes important
in the early and late era. This is an interesting solution with a new
dilaton potential which is shown in fig. 8. The late acceleration of
the universe may be explained in this framework quite successfully as
is shown in fig. 10 by the distance modulus curve. We also note cosmological
solution where the dilaton field behaves like phantom \cite{as:7}
in the early era but it transits to non-minimally coupled scalar field
in the later epoch in all the cases except in one case, where $\phi = H$
and $f(\phi)$ increases linearly with time. However it is evident from fig. 6 that there is a regim where $V(\phi)$ remains negative. We compare our Emergent
Universe model recent SNeIa data \cite{kn:1} at late universe. Fig. 10 show a comparative study of $\mu(z)$ vs. $z$ curve obtained from observation and that obtained theoretically from Emergent universe model. It may be pointed out here that between an exponential and accelerating phase, there might exist a phase of
particle creation era which is pointed out in {\it Ref.} \cite{kn:24} in the context of EU. However, a detail study on the issue will be taken up elsewhere.

\section*{Acknowledgments} BCP would like to thank Prof. S. Randjbar-Daemi for supporting  ICTP visit where this work was initiated and {\bf Third World Academy of Sciences (TWAS)} for awarding visiting Associateship to visit Institute of Theoretical Physics, Beijing Peoples Republic of China.  BCP would also like to thank UGC for awarding  Minor Research Project grant (No. F. 32-63/2006 (SR)). SG likes to thank {\it IUCAA Reference Centre } at  North Bengal University for extending facilities to do research work and the  University of  North Bengal for awarding Junior and Senior Research Fellowship. The authors are thankful to the referee for his constructive suggestions which helped in improving the presentation of the paper.

\end{document}